\documentclass[aps,pre,twocolumn,amsmath,showpacs,superscriptaddress]{revtex4}

\usepackage{amsmath}
\usepackage{amsbsy}
\usepackage{amscd}
\usepackage{amsopn}
\usepackage{amstext}
\usepackage{amsxtra}
\usepackage{epsfig}

\def\kp{{k^\prime}}

\def\lesssim{\mathrel{\hbox{\rlap{\hbox{\lower4pt\hbox{$\sim$}}}\hbox{$<$}}}}

\def\a{\alpha}

\def\be{\begin{equation}}

\begin{document}
\title{How clustering affects the bond percolation threshold in complex networks}
\author{ James P. Gleeson}
\affiliation{Department of Mathematics \& Statistics, University
of Limerick, Ireland.}
\author{Sergey Melnik}
\affiliation{Department of Mathematics \& Statistics, University
of Limerick, Ireland.}
\author{Adam Hackett}
\affiliation{Department of Mathematics \& Statistics, University
of Limerick, Ireland.}
\date{15 Apr 2010} 

\pacs{89.75.Hc, 64.60.aq, 64.60.ah, 87.23.Ge}

\begin{abstract}

The question of how clustering (non-zero density of triangles) in
networks affects their bond percolation threshold has important
applications in a variety of disciplines. Recent advances in
modelling highly-clustered networks are employed here to
analytically study the bond percolation threshold. In comparison
to the threshold in an unclustered network with the same degree
distribution and correlation structure, the presence of triangles
in these model networks is shown to lead to a larger bond
percolation threshold (i.e. clustering \emph{increases} the
epidemic threshold or \emph{decreases} resilience of the network
to random edge deletion).
\end{abstract}
 \maketitle
\section{Introduction}\label{sect:introduction}
Clustering (or \emph{transitivity}) in a complex network refers to
the propensity of two neighbors of a given node to also be
neighbors of each other, thus forming a triangle of edges within
the graph. In a recent paper \cite{Newman09a} Newman proposes a
model of random networks with clustering which permits analytical
solution for many important properties. An alternative model,
based on embedding cliques in a locally tree-like structure, was
subsequently proposed by one of us \cite{Gleeson09a}. One of the
most important predictions of these models is the effect of
network clustering on the bond percolation process, which is a
topic of considerable interest \cite{Newman03b, Britton07,
Serrano06c, Serrano06a, Serrano06b, Eames08, Miller08, Kiss08,
Trapman07, Gleeson09b}.

The bond percolation problem for a network may be stated as
follows: each edge of the network graph is visited once, and
\emph{damaged} (deleted) with probability $1-p$. The quantity $p$
is the \emph{bond occupation probability} and the non-damaged
edges are termed \emph{occupied}. In an infinite graph, the size
of the \emph{giant connected component} (GCC) of the graph becomes
nonzero at some critical value of $p>0$: this critical value of
$p$ is termed the \emph{bond percolation threshold}, denoted
$p_{th}$. The bond percolation problem has applications in
epidemiology, where $p$ is related to the average transmissibility
of a disease and the GCC represents the size of an epidemic
outbreak \cite{Grassberger83,Newman02b}, and in the analysis of
technological networks, where the resilience of a network to the
random failure of links is quantified by the size of the
GCC~\cite{Serrano06b}. Analytical solutions for percolation on
randomly-wired
networks and on correlated networks are well-known
\cite{Molloy95,Cohen00,Callaway00,Newman01a,Vazquez03,Goltsev08},
but these cases have zero clustering in the limit of infinite
network size.

Newman solves the bond percolation problem within his
model~\cite{Newman09a} and considers the effect of clustering on
the bond percolation threshold. He gives an example where
clustering decreases the value of $p_{th}$ within the context of a
certain set of networks which all share the same average degree
(see Fig. 2 of \cite{Newman09a}). However, Newman notes that the
networks in his comparison set, while having the same average
degree, do not all have the same degree distribution (see
Section~\ref{sect3} for further discussion of this point). Miller
\cite{Miller09} recently showed analytically that within the model
\cite{Newman09a} the bond percolation threshold in a clustered
network is greater that the corresponding threshold in an
unclustered network with the same degree distribution and
correlation structure. A similar conclusion was reported by Kiss
and Green \cite{Kiss08} based on their numerical simulations using
Newman's clustered bipartite graph model \cite{Newman03b}. In this
paper we focus on  networks generated by the clique-based model
\cite{Gleeson09a} and show that the effect of clustering is
qualitatively similar to that determined by Miller for the
triangle-based model \cite{Newman09a}, i.e., the presence of
clustering \emph{increases} the bond percolation threshold (and
hence the epidemic threshold) when networks with the same degree
distribution and correlation structure are compared. We emphasize
that the degree-degree correlation structure in the clustered
network includes non-trivial correlations beyond
nearest-neighbors, and we consider the implications of this fact.

  We begin by introducing the recently published models for
  clustered random networks, and in Section~\ref{sect2} we apply
  these to random regular graphs. Networks with heterogeneous
  degree distributions are examined in Sections~\ref{sect3} and
  \ref{sect4}, and conclusions are drawn in Section~\ref{sect5}.
  Extended mathematical calculations are relegated to the appendices.

We first briefly review two recent models for infinite random
networks with non-zero clustering. The fundamental quantity
describing the networks of \cite{Gleeson09a} is the joint
probability distribution $\gamma(k,c)$, giving the probability
that a randomly-chosen node has degree $k$ and is a member of a
$c$-clique (a fully-connected subgraph of $c$ nodes). In these
networks, nodes may be part of at most one clique. Nodes which are
members of a $c$-clique have $c-1$ edges linking them to neighbors
within the same clique. They also have  an additional $k-(c-1)$
neighbors who are not in the same clique as themselves (note
$\gamma(k,c)=0$ for $c>k+1$ since nodes in a $c$-clique must have
at least $c-1$ neighbors). Edges which are not internal to a
clique are termed \emph{external links}. The degree distribution
$P_k$ of the network (probability that a random node has $k$
neighbors) is obtained from $\gamma$ by averaging over all
possible clique sizes:
\begin{equation}
P_k = \sum_{c=1}^{k+1} \gamma(k,c) = \sum_c \gamma(k,c) \label{1}
\end{equation}
and
the degree-dependent clustering coefficient $c_k$ \footnote{
 The \emph{local clustering
coefficient} for a node $A$ is defined as the fraction of pairs of
neighbors of node $A$ which are also neighbors of each
other~\cite{Watts98}, and the \emph{degree-dependent clustering}
$c_k$ is the average of the local clustering coefficient over the
class of all nodes of degree $k$~\cite{Serrano06a,Vazquez02}.} is
given in terms of $\gamma$ by
\begin{equation}
c_k = \sum_c \frac{\gamma(k,c)}{P_k} \frac{(c-1)(c-2)}{k(k-1)},
\label{2}
\end{equation}
see \cite{Gleeson09a} for details. The overall network clustering
coefficient $C$~\cite{Watts98} is then $C=\sum_{k \ge 2}P_k c_k$.

Analytical results for  the giant connected component size are
given in \cite{Gleeson09a} and the bond percolation threshold
$p_{th}^{(\gamma)}$ is shown to be the solution of the following
polynomial equation for $p$:
\begin{equation}
 \frac{1}{z_e}\sum_{k,c} (k-c+1) \gamma(k,c)\left(p(k-c) +
(z_c-c+1)D_c(p)\right) = 1 \label{7}.
\end{equation}
Here $z_e$ is the average number of external links per node: $z_e
= \sum_{k,c}(k-c+1)\gamma(k,c)$, $z_c$ is the average degree of
nodes in cliques of size $c$: $z_c = \sum_k k \gamma(k,c)
/\sum_{k} \gamma(k,c)$, and $D_c(p) = p \sum_{m=1}^c (m-1) P(m|c)$
are polynomial functions of $p$. The functions $P(m|c)$ give the
probability that a node in a $c$-clique belongs to a connected
cluster of $m$ nodes within the clique, including itself; these
polynomial functions of $p$ are defined and tabulated in
\cite{Newman03b}.

A different approach to modelling local clustering is taken in
Newman's model \cite{Newman09a} (see also \cite{Miller09}). The
joint distribution $p_{s,t}$ gives the probability that a
randomly-chosen node is connected to $s$ single edges (similar to
the external links of the $\gamma$-theory networks) and to $t$
triangles. The degree distribution is then given by
\begin{equation}
P_k = \sum_{s,t}p_{s,t} \delta_{k,s+2t}
\end{equation}
and the clustering coefficient, GCC size, and bond percolation
threshold (denoted $p_{th}^{(N)}$ for Newman's model) may all be
determined analytically (see \cite{Newman09a,Miller09} and
Appendix~\ref{appA}).

It is instructive to compare the constraints imposed on the
network structure in each of these models. In Newman's model, a
$k$-degree node may be a member of up to $\lfloor k/2 \rfloor$
disjoint triangles, and thus have a local clustering coefficient
of up to $1/(k-1)$ if $k$ is even, or up to $1/k$ if $k$ is odd.
In contrast, nodes in the $\gamma$-theory networks can be members
of only a single clique, but using large cliques can give
arbitrarily high clustering. In Section~\ref{sect2} we show that
both models imply  $p_{th}$ is increased by clustering on random
regular graphs---this has recently been demonstrated for the case
of triangle-based networks \cite{Newman09a} by Miller
\cite{Miller09}, but we focus on the case of higher-clustering
$\gamma$-theory networks. A special class of clustered networks
are those whose nodes may belong to at most one triangle. Both
models \cite{Newman09a,Gleeson09a} are applicable to networks in
this class, and in Section~\ref{sect3} (see, for example,
Fig.~\ref{fig3}) we illustrate the interaction between clustering
and correlation common to both models of clustering.

\section{Random regular graphs} \label{sect2}
In this Section we restrict our attention to random $z$-regular
graphs, i.e., random graphs in which all nodes have the same
degree $z$. As shown in \cite{Newman01a} random graphs with zero
clustering (in the limit $N\to \infty$ of infinite number of
nodes) may be generated using the configuration model
\cite{Bender78,Bollobas80}, for which the percolation threshold is
given in terms of the degree distribution $P_k$ as
\begin{equation}
p_{th}^{(1)} = \frac{\sum_k k P_k}{\sum_k k(k-1)P_k }.\label{p1}
\end{equation}
 For random regular graphs the degree distribution is simply $P_k
= \delta_{k,z}$, and the zero-clustering percolation threshold is
$p_{th}^{(1)}= \frac{1}{z-1}$.

Next we employ  Eq.~(\ref{7}) to consider the effect of non-zero
clustering in regular networks generated using the algorithm of
\cite{Gleeson09a}. In \cite{Gleeson09a}  a parametrization of
$\gamma(k,c)$ is suggested which is consistent with (\ref{1}) and
allows the clustering to be easily adjusted:
\begin{equation}
\gamma(k,c) = P_k \,\left(\!\! \begin{array}{c} k\\c-1
\end{array}\!\! \right) g_k^{c-1}(1-g_k)^{k-c+1} \label{8}.
\end{equation}
 This is a binomial distribution of the
probability mass for $k$-degree nodes across the $c$-clique
classes for $c$ from $1$ to $k+1$, governed by the parameter
$g_k$. Substituting (\ref{8}) into (\ref{2}) gives the remarkably
simple relation $c_k=g_k^2$ between the degree-dependent
clustering coefficient and the parameter $g_k$. For the random
regular graphs under consideration here, $\gamma(k,c)$ is nonzero
only for $k=z$ and setting $g_z=\sqrt{C}$ in (\ref{8}) allows us
to investigate regular graphs with clustering coefficient $C$
covering the full range $[0,1]$.

\begin{figure}
\centering \epsfig{figure=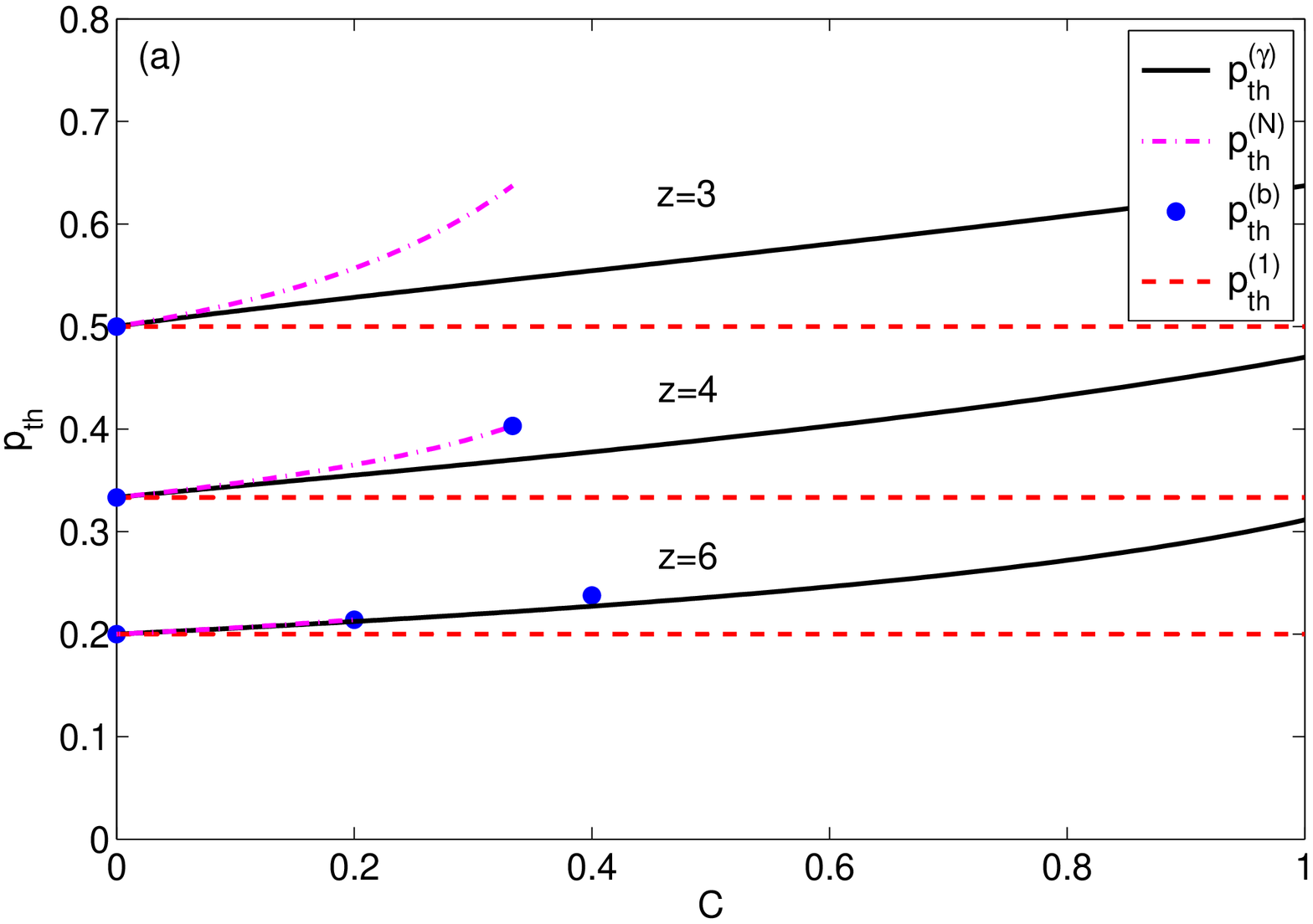,width=9cm}
\epsfig{figure=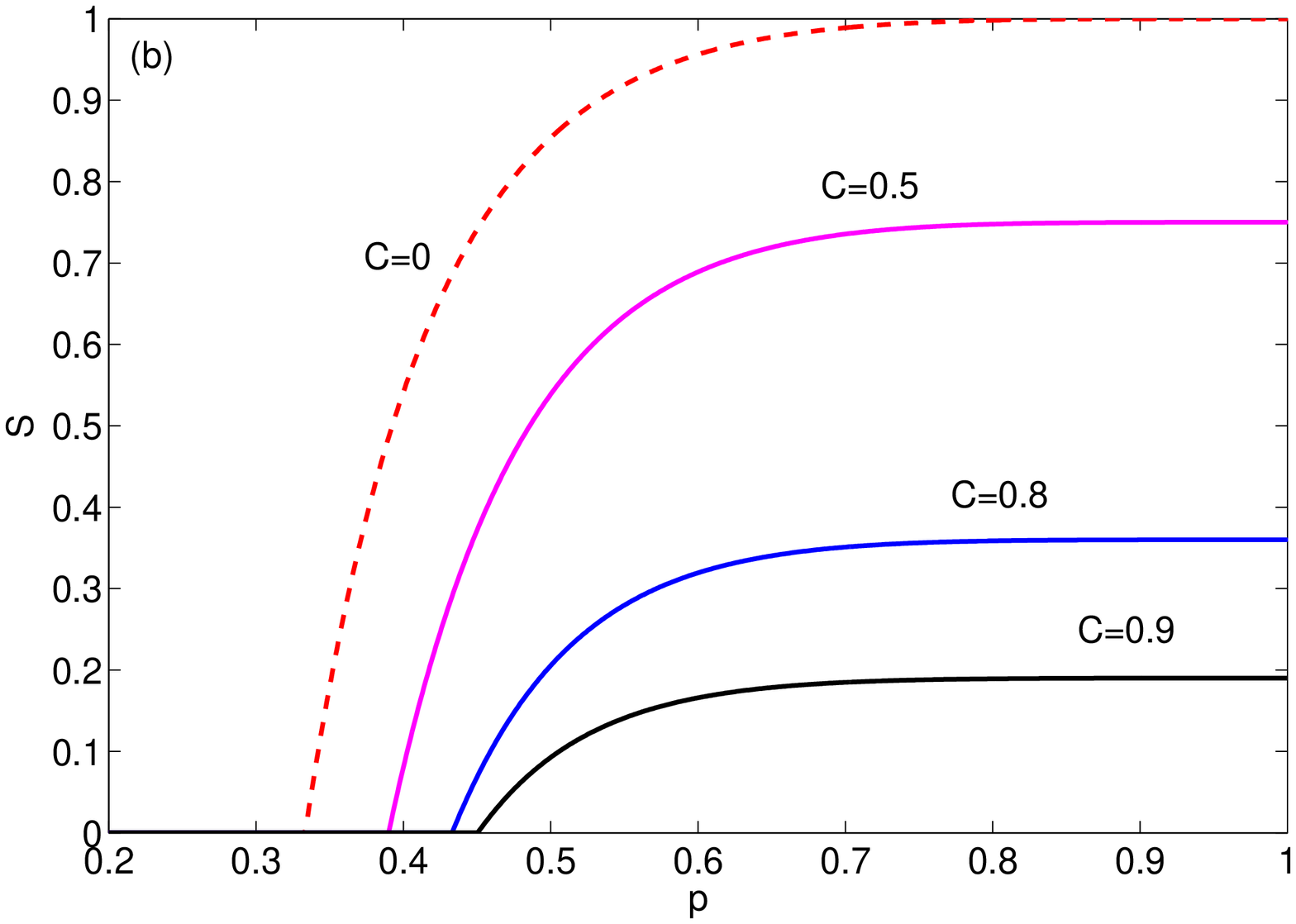,width=9cm}
 \caption{ (Color online) (a) Bond percolation threshold in
  $z$-regular graphs with clustering $C$, generated using the
   algorithms of \cite{Gleeson09a} ($p_{th}^{(\gamma)}$, black solid), \cite{Newman09a}
    ($p_{th}^{(N)}$, magenta dash-dot), and \cite{Newman03b} ($p_{th}^{(b)}$, blue symbols). For
    comparison, the threshold $p_{th}^{(1)}$ in an unclustered $z$-regular
    graph is shown by the red dashed line. Note $p_{th}^{(\gamma)}=p_{th}^{(N)}=p_{th}^{(b)}=p_{th}^{(1)}$ when $C=0$, but
    the clustered cases all have $p_{th}$ values exceeding $p_{th}^{(1)}$ when $C>0$. Values of $z$ are $z=3$ (top), $z=4$ (middle), and $z=6$ (bottom).
    (b) Sizes of GCC $S(p)$ in $z=4$ $\gamma$-theory regular graphs with clustering coefficients as shown.} \label{fig1}
\end{figure}

Figure~\ref{fig1}(a) compares the bond percolation threshold
$p_{th}^{(\gamma)}$ in clustered $\gamma$-theory networks
(determined by numerical solution of the polynomial Eq.~(\ref{7}),
using parametrization (\ref{8})) with the zero-clustering
threshold $p_{th}^{(1)}=1/(z-1)$. We also show (magenta dash-dot
curves) the percolation threshold $p_{th}^{(N)}$ given by Newman's
model \cite{Newman09a}, and the symbols show the threshold
$p_{th}^{(b)}$ found from an earlier bipartite-graph model of
clustering \cite{Newman03b}, see Appendix~\ref{appA} for details.
It is clear that all three clustering models give thresholds which
are larger than $p_{th}^{(1)}$ for $C>0$, i.e., \emph{clustering
increases the bond percolation threshold in these random regular
graphs}. Support for this statement in the case of $\gamma$-theory
networks is given in Appendix~\ref{appB}. The corresponding result
for $p_{th}^{(N)}$ follows from the recent work of Miller
\cite{Miller09}.

Analytical expressions determining the size $S$ of the giant
connected component in $\gamma$-theory networks are also given in
\cite{Gleeson09a} and Fig.~\ref{fig1}(b) shows $S$ as a function
of bond occupation probability $p$ for $z=4$, using
parametrization (\ref{8}). As already noted, increased clustering
leads to higher values of the transition point
$p_{th}^{(\gamma)}$, but also leads to smaller GCC sizes.

Having  established that the presence of clustering increases
$p_{th}$ in several models of clustered  regular graphs, in the
remainder of this paper we will consider how diversity of node
degrees also plays an important role.

\section{Heterogeneous networks} \label{sect3}

Networks with a range of node degrees may be characterized at
first order by their degree distribution $P_k$ or, at second
order, by the joint probability $P(k,\kp)$ that a randomly-chosen
edge links vertices of degree $k$ and $\kp$. Analytical results
for the percolation threshold are known for the ensembles of
networks described fully by $P_k$ \cite{Callaway00} or by
$P(k,\kp)$ \cite{Vazquez03} with respective thresholds denoted
$p_{th}^{(1)}$ and $p_{th}^{(2)}$, see (\ref{p1}) and
Appendix~\ref{appC}.

\begin{figure}
\centering \epsfig{figure=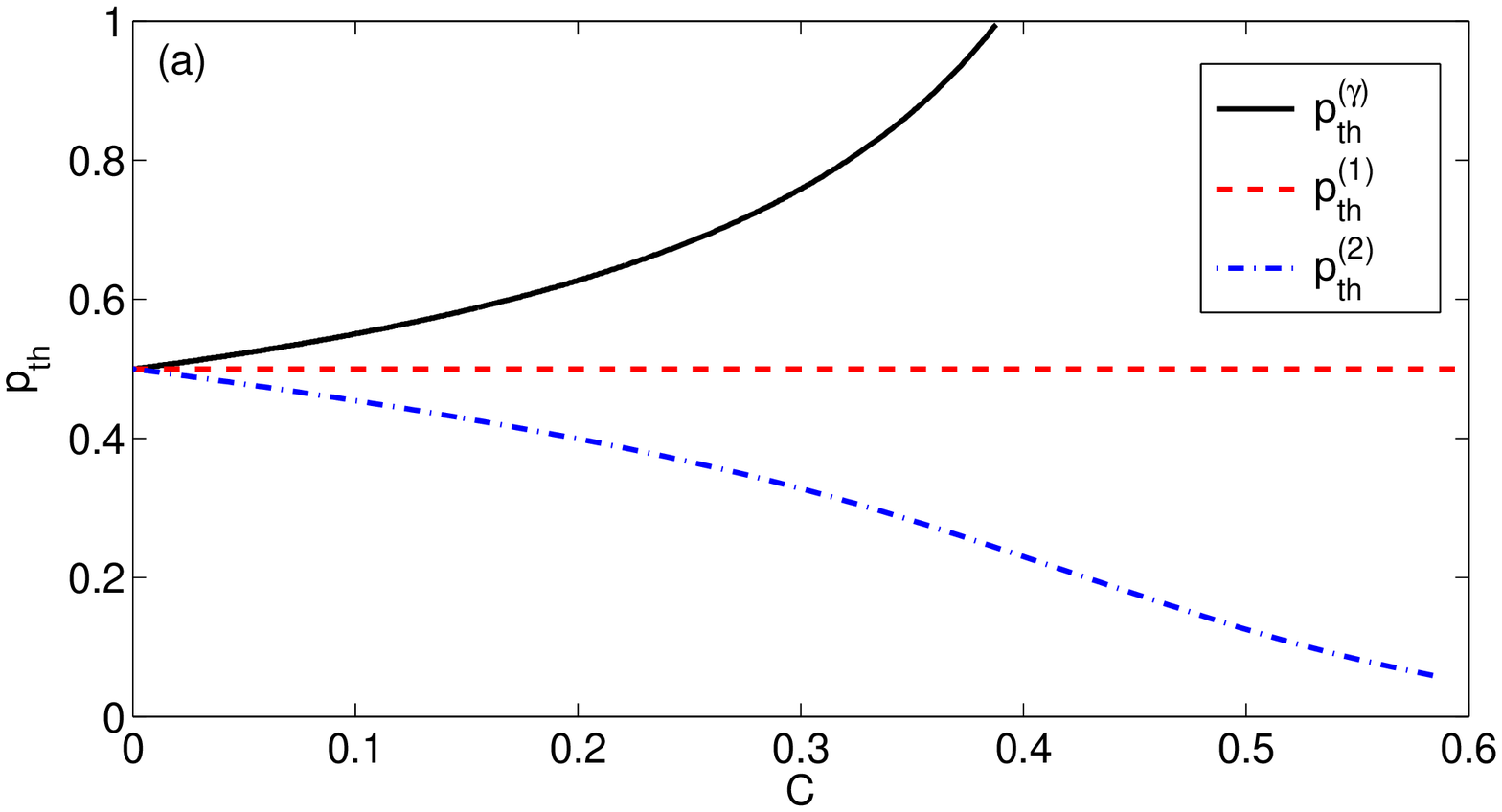,width=9cm}
\epsfig{figure=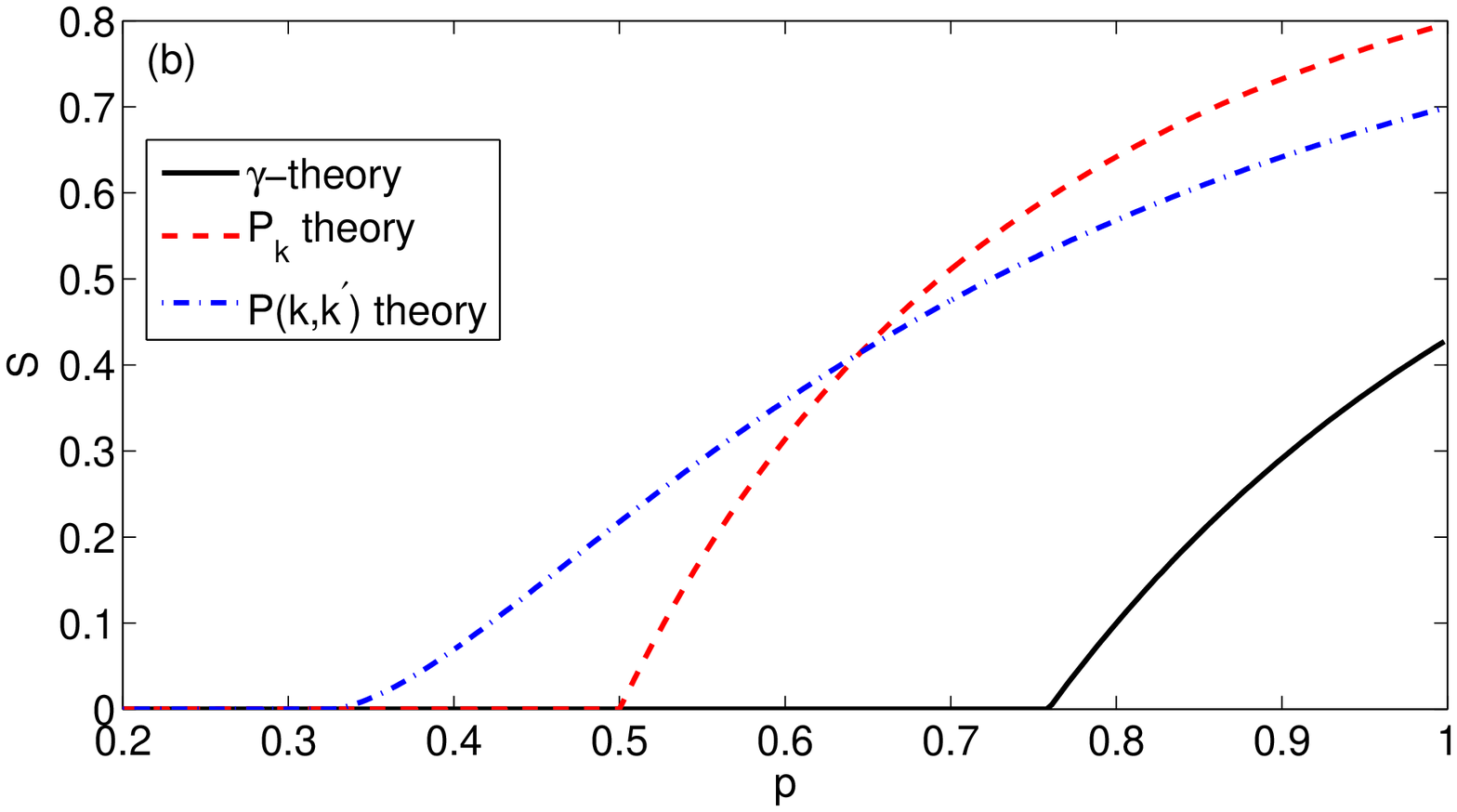,width=9cm}
 \caption{ (Color online) (a) Bond percolation threshold in $\gamma$-theory networks with
 Poisson degree distribution, $z=2$, and clustering $C$ (black solid). For comparison, also shown
 is the threshold $p_{th}^{(1)}$ in an unclustered network with same degree distribution (red dashed), and the
 threshold $p_{th}^{(2)}$ in an unclustered network with the same degree-degree correlations (blue dash-dot) as the $\gamma$-network.
  (b) Sizes of GCC $S(p)$
 for the case $C=0.3$ in $\gamma$-theory networks (black solid), and in unclustered
 networks with the same degree distribution (red dashed), or same degree-degree correlations (blue dash-dot).}
 \label{fig2}
\end{figure}

In this section we compare the bond percolation threshold
$p_{th}^{(\gamma)}$ for various clustered networks with the values
$p_{th}^{(1)}$ and $p_{th}^{(2)}$ corresponding to zero-clustering
networks with the same degree distribution, or same degree-degree
correlations as the clustered network. Our first example is a
Poisson random network with degree distribution $P_k = e^{-z}
z^k/k!$ and mean degree $z=2$. Figure~\ref{fig2}(a) compares
$p_{th}^{(\gamma)}$ from Eq.~(\ref{7}) with $p_{th}^{(1)}=1/z$ and
$p_{th}^{(2)}$, the latter being determined using the joint
distribution $P(k,\kp)$ for $\gamma$-theory networks derived in
Appendix~\ref{appC}. The clustering level of the $\gamma$-theory
networks is controlled using the parametrization (\ref{8}), with
$g_k = \sqrt{C/(1-P_0-P_1)}$ for all $k$, so that the average
clustering coefficient $\sum_{k\ge 2} P_k c_k$ is equal to $C$.
Note that the $p_{th}^{(1)}$ line (and $p_{th}^{(2)}$ curve) show
the thresholds in unclustered networks with the same degree
distribution (and $P(k,\kp)$ distribution) as the $\gamma$-theory
network with clustering $C$.

We see that $p_{th}^{(\gamma)}$ is larger than both of the
zero-clustering thresholds $p_{th}^{(1)}$ and $p_{th}^{(2)}$,
consistent with our claim that clustering increases the bond
percolation threshold. The fact that $p_{th}^{(2)}$ is less than
$p_{th}^{(1)}$ is due to the assortativity of the $\gamma$-theory
networks, see Appendix~\ref{appC} and \cite{Goltsev08}.

Figure~\ref{fig2}(b) shows the GCC size $S$ in the $\gamma$-theory
network (black solid curve) as a function of $p$ for clustering
$C=0.3$. Also shown are the GCC sizes in a zero-clustering network
with the same degree distribution $P_k$ (red dashed curve) and
with the same $P(k,\kp)$  distribution (blue dash-dot curve). This
figure can be compared to Fig.~2 of \cite{Newman09a} where
higher-clustering cases seem to have lower percolation thresholds
than the zero-clustering case.
  However,
it should be noted that the focus in \cite{Newman09a} is on a
different comparison to that undertaken here. The cases plotted in
Fig.~2 of \cite{Newman09a} are  generated from a double Poisson
$p_{s,t}$
  distribution (see Eq.~(13) of \cite{Newman09a}) and all
  share the same mean degree $z$, but not the same degree
  distribution.
  In short, we
  compare clustered networks with unclustered versions with the
  same $P_k$ (or $P(k,\kp)$), while Newman's comparison in \cite{Newman09a} retains a
  common form for the joint distribution $p_{s,t}$, but does not conserve the
  degree distribution. A similar analysis applies to Fig.~2 of
  \cite{Newman03b}, where again it may be shown that the clustered
  networks used have percolation thresholds larger than those of
  unclustered networks with the same degree distribution. In fact
  this has been demonstrated numerically by Kiss and Green \cite{Kiss08},
  who
  compared the GCC sizes for the networks of \cite{Newman03b}
  with the GCC sizes in rewired versions of these networks.

\begin{figure}
\centering
\epsfig{figure=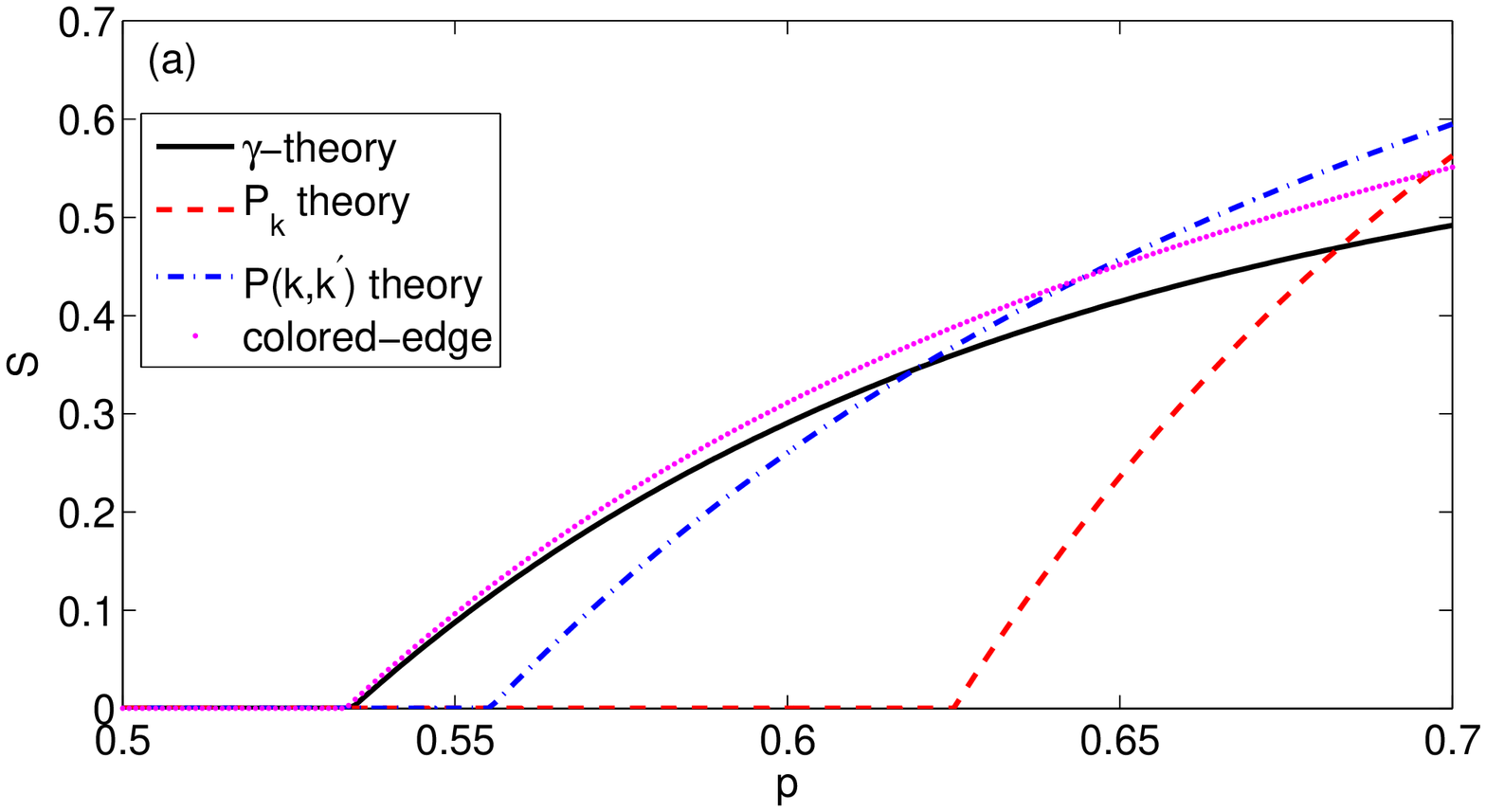,width=8.2cm}
\epsfig{figure=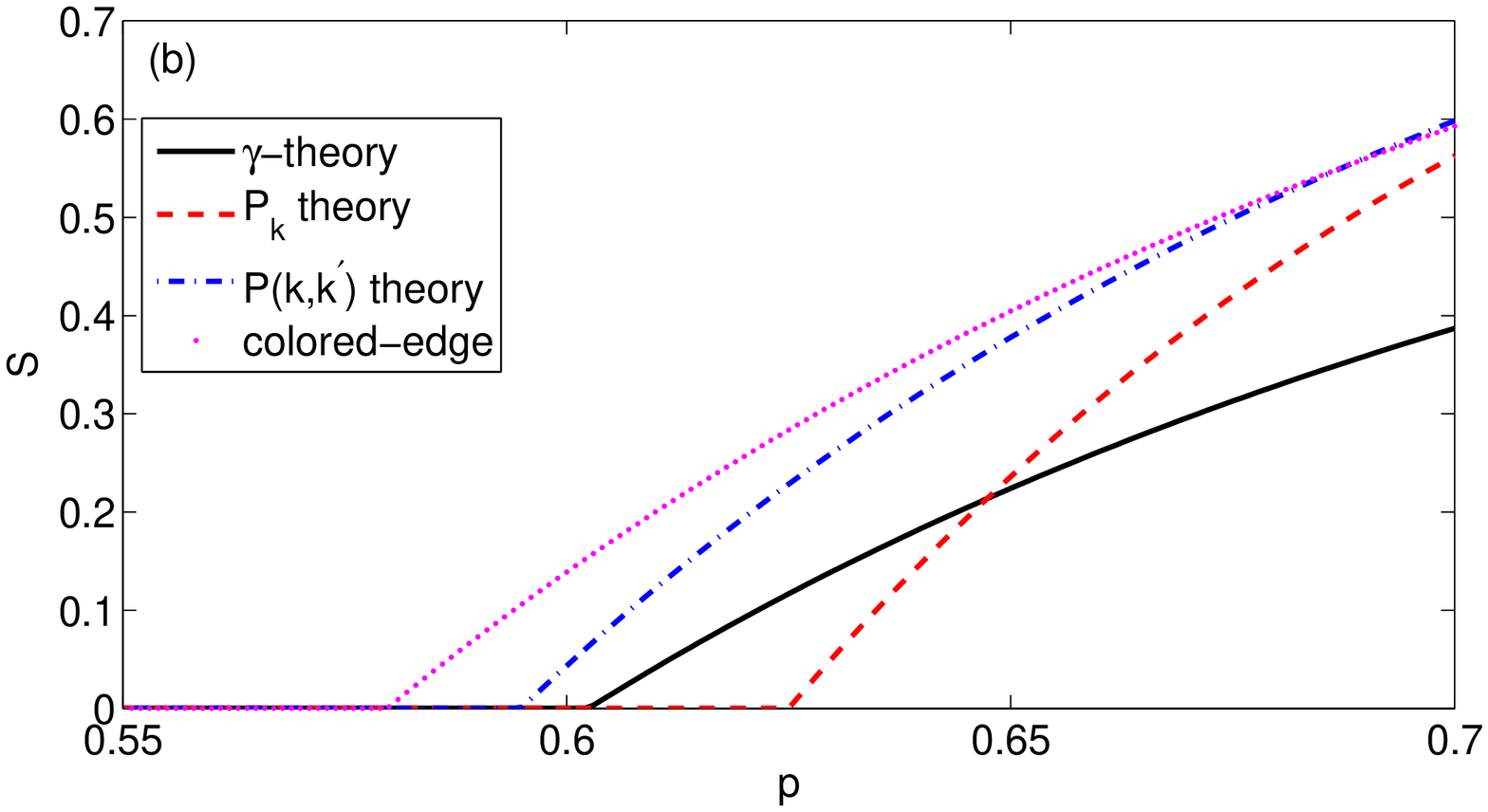,width=8.2cm}
\epsfig{figure=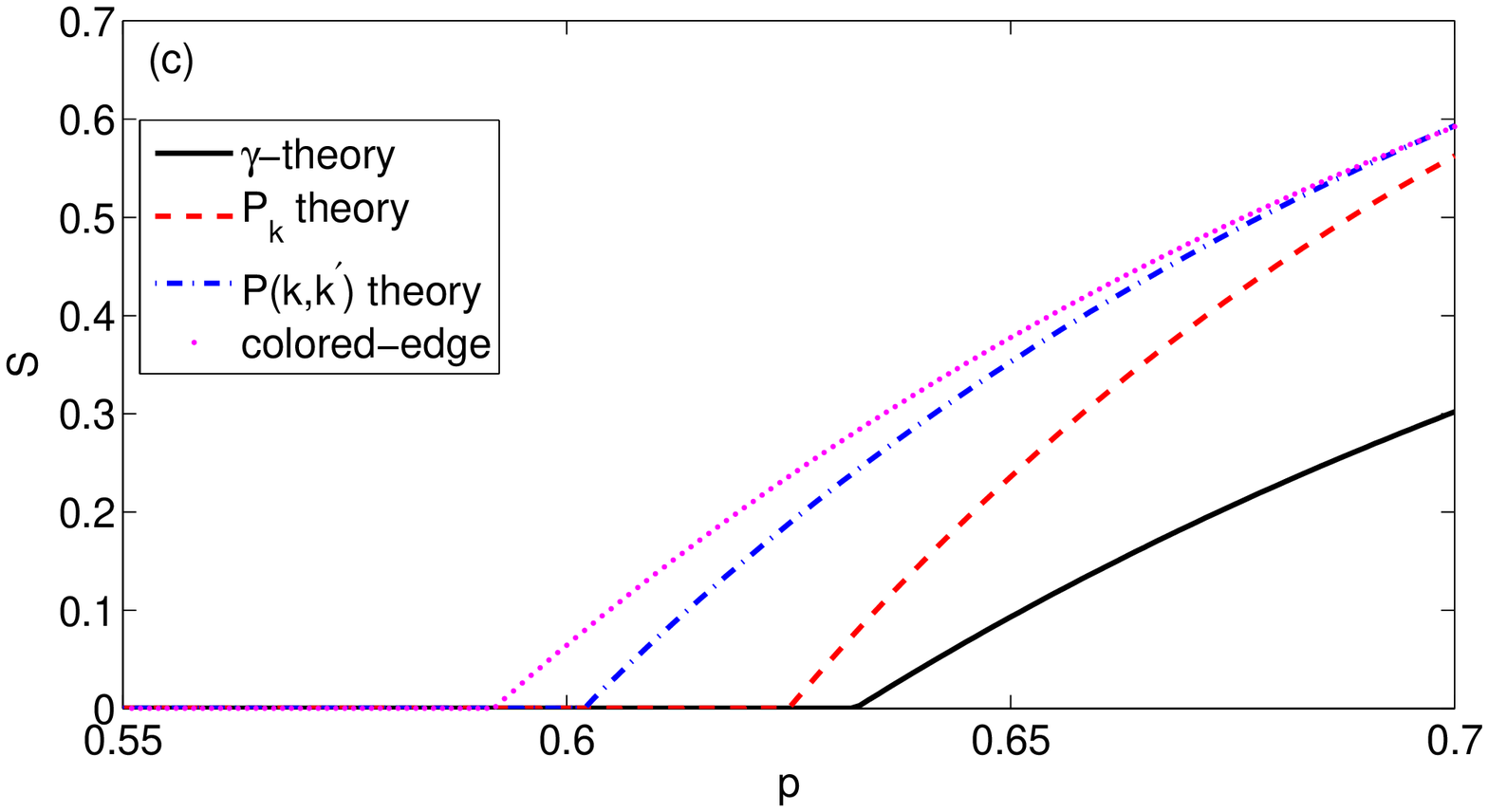,width=8.2cm}
 \caption{ (Color online) Sizes of GCC $S(p)$ for the
  $\gamma$-theory networks defined
 by (\ref{gamma4}) (black solid) and in unclustered
 networks with the same degree distribution (red dashed), or same degree-degree correlations (blue dash-dot). The magenta
 dotted curve is for the colored-edge (unclustered) networks defined in Section~\ref{sect4}.
 Parameters are $\alpha=0.9$, with (a) $\beta=0.1$, (b) $\beta=0.4$, and (c) $\beta=0.5$.} \label{fig3}
\end{figure}

  Having examined the results for regular graphs and Poisson
  random networks, one might be tempted at this point to conclude
  that $p_{th}^{(\gamma)}$ is always greater than $p_{th}^{(1)}$
  and $p_{th}^{(2)}$. However, the situation is rather more
  complicated than this, as demonstrated in Fig.~\ref{fig3} and discussed (for Newman's triangle-based networks)  in \cite{Miller09}. To
  facilitate analysis, and to enable the application of both the
  $\gamma$-theory \cite{Gleeson09a} and Newman's theory
  \cite{Newman09a}, we restrict our attention now to the
  special class of
  networks in which each node has either zero local clustering, or
  is part of a single triangle. In terms of the $\gamma$-theory, this
  means $\gamma(k,c)=0$ unless $c=1$ or $c=3$. For Fig.~\ref{fig3}
  we have also used a particularly simple degree distribution,
  with exactly half the nodes having degree $k=2$ and the other
  half having degree $k=3$. The networks examined are thus
  described with the theoretical models as follows
  \begin{eqnarray}
  \gamma(2,1) &=& p_{2,0}=\frac{1}{2}(1-\alpha); \quad   \gamma(2,3)
  =
  p_{0,1}=\frac{1}{2}\alpha,\nonumber\\
 \gamma(3,1) &=& p_{3,0}=\frac{1}{2}(1-\beta); \quad  \gamma(3,3)
 =
  p_{1,1}=\frac{1}{2}\beta, \label{gamma4}
  \end{eqnarray}
  with the parameters $\alpha$ and $\beta$ controlling the
  level of clustering for each degree class.

  Figure~\ref{fig3} shows that $p_{th}^{(\gamma)}$ (which equals
  $p_{th}^{(N)}$ in this special class of networks) may lie either below (Fig.~\ref{fig3}(a)) or above (Fig.~\ref{fig3}(c))
   the
  zero-clustering thresholds $p_{th}^{(1)}$ and $p_{th}^{(2)}$.
  Recall our claim is that the presence of triangles increases
  $p_{th}$ relative to its value in unclustered networks with the
  same degree distribution \emph{and  same correlation structure}. In
  the next section we show that the correlation structure in these
  examples is not fully described by only nearest-neighbor
  correlations as given by $P(k,\kp)$. When, as described in Section~\ref{sect4},  the correlation
  structure is fully matched but clustering eliminated, the GCC
  size $S(p)$ is given by the magenta (dotted) curve in Fig.~\ref{fig3}. Note
  the transition point for the black (solid) curve is larger in all cases
  than the transition point for the magenta curve, supporting our
  claim. Detailed analysis of the correlation structure for these
  cases is given in Section~\ref{sect4} and Appendix~\ref{appD}.

  \section{Unclustered networks with correlation structure} \label{sect4}

In this section we restrict our attention to the special class of
$\gamma$-theory networks wherein nodes are members of either one
clique or of none, and all cliques are of equal size $c=\overline
c$ (the example in Section~\ref{sect3} used $\overline c=3$),
i.e.,
\begin{equation}
\gamma(k,c) = P_k(1-\a_k)\delta_{c 1} + P_k \a_k \delta_{c
\overline c},
\end{equation}
for a prescribed degree distribution $P_k$, and with $\alpha_k$
determining the level of clustering for degree-$k$ nodes. Note
that the theoretical approaches of \cite{Newman09a} and
\cite{Gleeson09a} both apply in the case $\overline c =3$.

\begin{figure}
\centering \epsfig{figure=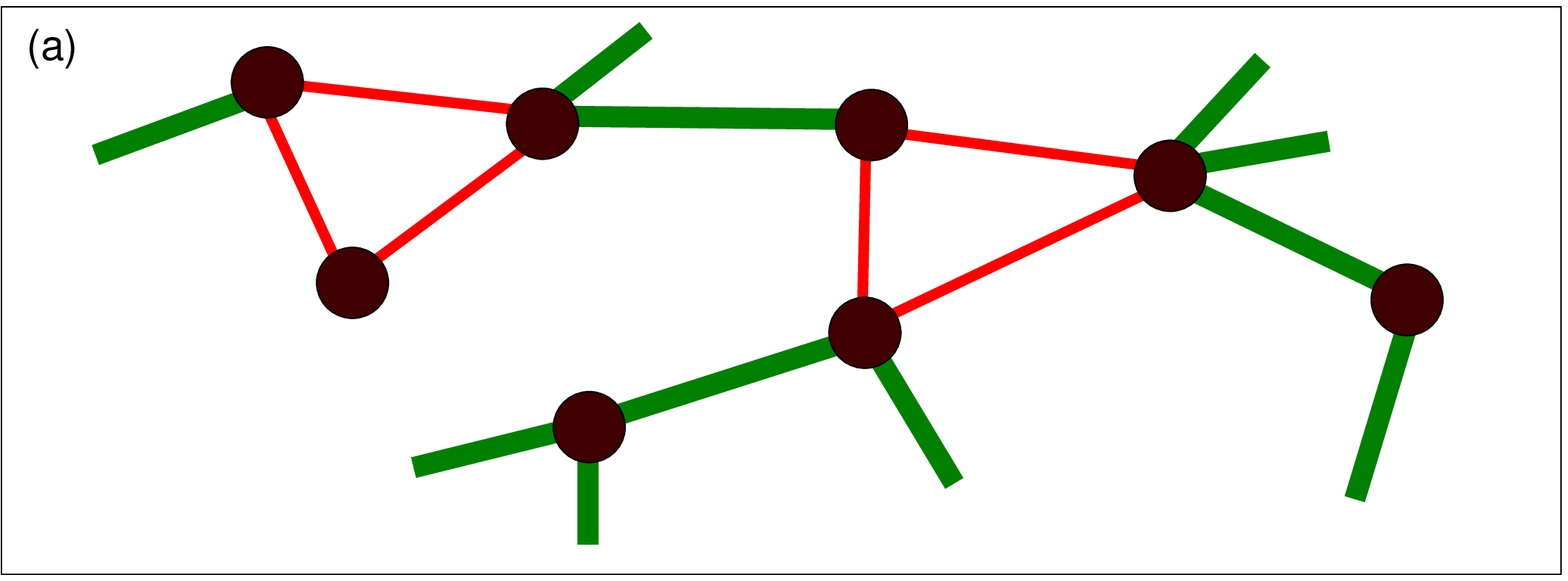,width=9cm}
\epsfig{figure=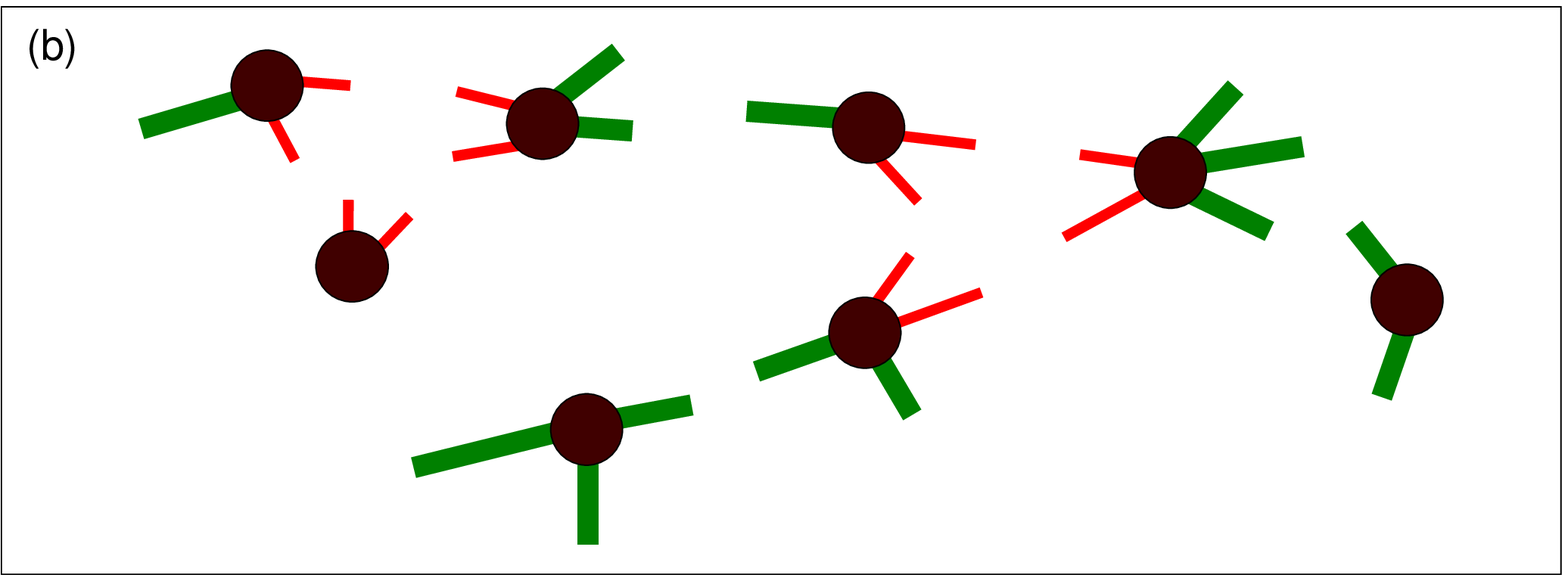,width=9cm}
 \caption{ (Color online) Segment of a clustered network with clique edges colored red (thin lines) and
 external links colored green (thick lines). After breaking each edge to obtain colored stubs as in (b), a realization
 of a colored-edge network is created by randomly connecting pairs of stubs of the same color.}
 \label{fig4}
\end{figure}

To understand the correlation structure of these networks we
visualize each edge of a network as being colored either green or
red (compare to the approach for the triangle-based Newman model
taken recently in \cite{Miller09}). The rule for edge-coloring is
simple: all edges which form part of a $\overline c$-clique  are
colored red, while the remaining edges (the external links in the
$\gamma$-theory notation) are all colored green, see
Fig.~\ref{fig4}(a) for an example with $\overline c = 3$. Now
consider the following rewiring process, which preserves the
correlation structure, but destroys the clustering within the
network. First, break each edge into two end-stubs with each stub
retaining the color of the original edge. We now have $N$ isolated
``hedgehog'' nodes, each with a set of colored stubs as its
``spines'', see Fig.~\ref{fig4}(b). The network is then
reconnected together by randomly selecting pairs of green stubs to
be joined with a green edge, and similarly randomly pairing red
stubs with red edges. The construction method for the original
$\gamma$-theory (or Newman theory) involves a similar joining of
like-colored stubs, except that the randomly chosen red stubs are
gathered into $\overline c$-cliques. By simply joining pairs of
red stubs at random we retain the degree-degree correlation
structure (including correlations beyond nearest-neighbor) of the
$\gamma$-theory network, but eliminate triangles (in the $N \to
\infty$ limit). The resulting network, which we dub the
\emph{colored-edge network}, has properties which are influenced
by the fact that red and green stubs are not randomly distributed
among the nodes. Taking $\overline c =3$ for example, each node is
a  member of 0 or 1 triangle, so we know that each node must have
either exactly zero or exactly two red edges linked to it, while a
node of degree $k$ has either $k$ or $k-2$ green edges. These
constraints mean the correlation structure of the colored-edge
network is not completely described only by the nearest-neighbor
correlations (i.e., by the $P(k,\kp)$ distribution of
Appendix~\ref{appC}). A worked example showing this
 correlation structure is given in
Appendix~\ref{newApp}.

Despite the non-trivial correlation structure, the lack of
clustering permits the application of standard tree-based
approaches to find the GCC size and the bond percolation threshold
$p_{th}^{(ce)}$ for colored-edge networks generated from
$\gamma$-theory networks with the single non-trivial clique class
$c=\overline c$ (see \cite{Miller09} for the case $\overline c=3$,
and Appendix~\ref{appD} for the general $\overline c$ case). The
magenta dotted curve in Fig.~\ref{fig3} shows the GCC size for the
colored-edge networks. In Appendix~\ref{appD} we show analytically
that $p_{th}^{(ce)} \le p_{th}^{(\gamma)}$, i.e., that the
clustering in the original network causes it to have an increased
bond percolation threshold compared to the colored-edge network
with the same correlation structure. However, the relative
ordering of $p_{th}^{(ce)}$ and $p_{th}^{(1)}$ (or
$p_{th}^{(2)}$)---and hence the ordering of $p_{th}^{(\gamma)}$,
$p_{th}^{(1)}$, $p_{th}^{(2)}$---depends on the details of the
correlation structure beyond nearest-neighbors, so the fact that
$p_{th}^{(\gamma)}$ exceeds $p_{th}^{(ce)}$ does not guarantee it
will exceed $p_{th}^{(2)}$, see Fig.~\ref{fig3}(a) for an example.
Further work is needed to elucidate the effects of the correlation
structure on $p_{th}$ in these unclustered networks, but we
believe the effect of clique-based clustering has now been clearly
separated from this question.

\section{Conclusions} \label{sect5}
We have shown that within the context of the clique-based model of
\cite{Gleeson09a}, clustering increases the bond percolation
threshold in comparison with its value for networks with (i) the
same degree distribution and (ii) the same correlation structure.
 In
Section~\ref{sect2} we used three different approaches for
constructing random regular networks with clustering, and
confirmed that $p_{th}$ is increased by the presence of
clustering, both in triangle-based networks (as shown in
\cite{Miller09}) and also in the highly-clustered clique-based
models of \cite{Newman03b} (as first demonstrated in
\cite{Kiss08}) and \cite{Gleeson09a} (see Fig.~\ref{fig1} and
 Appendix~\ref{appB}). In Sections \ref{sect3} and
\ref{sect4} we highlighted the importance of condition (ii) by
showing that the $n$th-nearest-neighbor correlations affect
$p_{th}$ even in the absence of clustering, i.e., networks with
identical nearest-neighbor correlations (as given by the
$P(k,\kp)$ distribution) can have differing $p_{th}$ due to
correlations beyond nearest neighbor. The $n$th-nearest-neighbor
correlations are therefore also important when investigating the
effects of clustering within various models. When these
correlations are fully accounted for, our result remains valid
(see Fig.~\ref{fig3} and Appendix~\ref{appD}).

What should be our intuitive understanding of the effects of
clustering? We believe the correct viewpoint was in fact given by
Newman \cite{Newman09a} when discussing the giant component size
in the case $p=1$: \emph{``the triangles that give the network its
clustering contain redundant edges that serve no purpose in
connecting the giant component together''}. In other words, the
redundant edges cause the GCC size in a clustered network to be
smaller than (or at most equal to) the GCC of an unclustered
network with the same correlation structure, thus explaining the
observation that
 clustering decreases the value of $S(1)$  in the Newman model \cite{Miller09}.
  All our results indicate that in fact
$S^{(\gamma)}(p) \le S^{(ce)}(p)$ for all $p$ in $[0,1]$, i.e.,
that clustering reduces the GCC size for all values of $p$
(compared, as usual, to an unclustered network with same
correlation structure), not just for $p=1$. Our main result, that
$p_{th}^{(\gamma)} \ge p_{th}^{(ce)}$, may be seen as a simple
consequence of this fact: since the GCC size in the clustered
network is smaller than (or at most equal to) that in the
unclustered network for all $p$, the transition point where the
clustered GCC size becomes nonzero must be larger than the
transition point for the unclustered network. We therefore believe
that Newman's explanation of clustering as adding redundant edges
reveals the essence of the matter.

In the recent paper \cite{Miller09}, Miller independently derives
the triangle-based clustering model of \cite{Newman09a}. He also
demonstrates that within the context of this model, clustering
increases the bond percolation threshold in the same sense as
claimed here (i.e., when compared to an unclustered network with
identical correlation structure). Our work is complementary to
\cite{Miller09}, since we show that the qualitative effect of
clustering seen in triangle-based networks (i.e. clustering
increases $p_{th}$) is also present in more heavily-clustered
networks described by clique-based theory (compare our results in
Appendices~\ref{appB} and \ref{appD} with those in
\cite{Miller09}).

The application of these results to real-world networks remains a
significant challenge. In this paper it was possible to separate
the effects of clustering and the related correlation structure
within the theoretical models \cite{Newman09a,Gleeson09a}, but it
is not clear how this might be attempted for a given real-world
network or indeed for other theoretical models with clustering.
Nevertheless, the understanding that within the models
\cite{Newman09a,Gleeson09a} clustering (as distinct from related
correlation effects) leads generically to an increase in the bond
percolation threshold marks, we believe, an important step
forward.

\section*{Acknowledgements}
This work was funded by Science Foundation Ireland under
programmes 06/IN.1/I366 and MACSI 06/MI/005.

\appendix
\section{Other clustering models}\label{appA}
 Newman's results
\cite{Newman09a} may be used to derive the following polynomial
equation for the bond percolation threshold $p=p_{th}^{(N)}$ in
networks described by the joint distribution $p_{s,t}$ (see also
\cite{Miller09}):
\begin{align}
\nonumber 2p(1+p-p^{2})(p(\langle s^{2}-s\rangle\langle t^{2}-t\rangle-\langle st\rangle^{2})-\langle s\rangle\langle t^{2}-t\rangle)\\
-p\langle s^{2}-s\rangle\langle t\rangle+\langle s\rangle\langle
t\rangle=0,
\end{align}
where $s$ and $t$  specify respectively the number of single edges
and triangle edges attached to a vertex, and $\langle\cdot\rangle$
denotes the average over the joint distribution $p_{s,t}$. For
random $z$-regular graphs we assume the following distribution of
probability mass:
\begin{equation}
p_{s,t}={\lfloor \frac{z}{2}\rfloor \choose t} g^{t}(1-g)^{\lfloor
\frac{z}{2}\rfloor-t}\quad \text{for $t=0$ to
$\left\lfloor\frac{z}{2}\right\rfloor$},
\end{equation}
 and calculate the
clustering $C$ in terms of the single parameter $g$ using the
results of \cite{Newman09a}. The magenta dash-dotted curves in
Fig.~\ref{fig1} show $p_{th}^{(N)}$ as a function of $C$.

Another analytically solvable case of clustered random regular
graphs is provided by Newman's bipartite graph model
\cite{Newman03b}. In this model, nodes may be part of some number
of groups (cliques), and the structure may be represented as a
bipartite graph with links between nodes (individuals) and the
groups (cliques) of which they are members. In general this model
cannot be fitted to desired degree distributions, but the special
case of $z$-regular graphs may be produced by taking the
distribution of group sizes to be $s_n = \delta_{n,\nu}$, and the
number of groups in which a node partakes to be distributed as
$r_m = \delta_{m, \mu}$, where integers $\nu$ and $\mu$ satisfy
the relation $(\nu-1)\mu = z$. For the case $z=6$, for example,
there exist 3 such $(\nu,\mu)$ pairs: $(2,6)$, $(3,3)$, and
$(4,2)$, leading to respective clustering coefficients of $0$,
$1/5$, and $2/5$. The formulas given in \cite{Newman03b} allow us
to calculate the bond percolation threshold for each of these
cases, and the results are plotted with symbols in
Fig.~\ref{fig1}(a). Consistent with the models of
\cite{Gleeson09a,Newman09a}, the percolation threshold is clearly
increased above its unclustered value in this model.

\section{Clustering increases $p_{th}$ in random regular graphs}
\label{appB} Here we demonstrate that for random $z$-regular
graphs generated using the $\gamma$-theory \cite{Gleeson09a}, the
bond percolation threshold $p_{th}^{(\gamma)}$ is larger than the
value $p_{th}^{(1)}=1/(z-1)$ for an unclustered network. We show
this for a general $\gamma(k,c)$ distribution, so the result is
not dependent on a particular parametrization such as (\ref{8}).

Note from (\ref{7}) that $p_{th}^{(\gamma)}$ is the solution of
the polynomial equation $F(p)=1$ where
\begin{eqnarray}
F(p) &=&
\frac{1}{z_e}\sum_{c}{(z-c+1)}\gamma(z,c)\times\nonumber\\
&& \hspace{0.2cm}\times\left(p(z-c) + (z-c+1)D_c(p)\right),
\label{B1}
\end{eqnarray}
with $z_e = \sum_c(z-c+1)\gamma(z,c)$. We use the following two
properties of the polynomials $D_c(p)$: (a) $D_c(p)$ is a
monotonically increasing function of $p$ on the interval $[0,1]$
with $D_c(0)=0$, and (b) $D_c(p)$ is bounded above  by
\begin{equation}
D_c(p) \le \frac{p^2(c-1)}{1-p(c-2)}\label{B2}
\end{equation}
for all $p$ with $0 \le p \le \frac{1}{c-2}$.

By property (a), the polynomial $F(p)$ defined in (\ref{B1}) is
monotonically increasing in $p$, with $F(0)=0$. Since
$F\left(p_{th}^{(\gamma)}\right)=1$, we can guarantee that
$p_{th}^{(1)} \le p_{th}^{(\gamma)}$ by showing that
$F\left(p_{th}^{(1)}\right) \le 1$. Using property (b), we have
that for $p\le \min_c \left(1/(c-2)\right)$,
\begin{eqnarray}
F(p)&\le&\frac{1}{z_e}\sum_c (z-c+1)\gamma(z,c) \times\nonumber\\
&& \times \left( p (z-c) + \frac{(z-c+1)p^2
(c-1)}{1-p(c-2)}\right).
\end{eqnarray}
Substituting $p=p_{th}^{(1)}=1/(z-1)$ (note this $p$ obeys $p \le
1/(c-2)$ for all relevant cliques classes since $c\le z+1$ in a
$z$-regular graph) simplifies the right-hand side to yield
\begin{eqnarray}
F\left(p_{th}^{(1)}\right) &\le& \frac{1}{z_e} \sum_c
(z-c+1)\gamma(z,c) \nonumber\\
&=& 1,
\end{eqnarray}
hence implying that $p_{th}^{(\gamma)}\ge p_{th}^{(1)}$ as
desired.

\section{Degree-degree correlations in $\gamma$-theory networks}\label{appC}
The ensemble of networks characterized by $\gamma(k,c)$ is
constructed as described in \cite{Gleeson09a}. To determine the
degree-correlation matrix $P(k,\kp)$ we calculate the probability
that a randomly-chosen edge of the network joins together nodes of
degree $k$ and $\kp$. The construction algorithm for the
$\gamma(k,c)$ network is based on specifying stubs (half-edges) as
either  \emph{external stubs} or \emph{$c$-clique stubs}. Since
each $k$-degree node in a $c$-clique has $k-c+1$ external stubs
and $c-1$ $c$-clique stubs, the number of external edges in the
network  (half the number of external stubs) is given by
\begin{equation}
E_e = \frac{N}{2} \sum_{k,c} (k-c+1)\gamma(k,c),
\end{equation}
where $N$ is the number of nodes. Similarly, the total number of
$c$-clique edges is
\begin{equation}
E_c = \frac{N}{2} \sum_k (c-1) \gamma(k,c), \text{ for } c>1.
\end{equation}
The sum over all $c$-clique classes, plus the external edges,
gives the total number $E$ of edges in the network:
\begin{equation}
E = E_e + \sum_{c > 1} E_c = \frac{1}{2}N z.
\end{equation}
Therefore a randomly-chosen edge of the network is an external
edge with probability $E_e/E \equiv \alpha^{(1)}$ and is a
$c$-clique edge with probability $E_c/E \equiv \alpha^{(c)}$. Then
the global $P(k,\kp)$ matrix may be written as
\begin{eqnarray}
P(k,\kp) &=& \frac{E_e}{E} P_e(k,\kp) + \sum_{c>1} \frac{E_c}{E}
P_c(k,\kp)\nonumber\\
&=& \alpha^{(1)} P_e(k,\kp) + \sum_{c > 1} \alpha^{(c)}
P_c(k,\kp), \label{new1}
\end{eqnarray}
where $P_e(k,\kp)$ is the probability that a randomly chosen
external edge joins nodes of degrees $k$ and $\kp$, and
$P_c(k,\kp)$ is similarly defined for $c$-cliques edges.

Suppose first that the chosen edge is an external edge. Since
external edges are composed of randomly-connected external stubs,
the probability that an end-vertex is of degree $k$ is
\begin{equation}
s_k^{(1)}=\sum_c \frac{(k-c+1)\gamma(k,c)}
{\sum_{k^{\prime},c^{\prime}}
(k^{\prime}-c^{\prime}+1)\gamma(k^{\prime},c^{\prime})}
\end{equation}
and the probability that the chosen external edge links nodes of
degrees $k$ and $\kp$ is
\begin{equation}
P_e(k,\kp) =  s_k^{(1)}s_\kp^{(1)}.\label{Pe}
\end{equation}
If the chosen edge is a $c$-clique edge, the probability that an
end-vertex is of degree $k$ is
\begin{equation}
s_k^{(c)} =
\frac{(c-1)\gamma(k,c)}{\sum_{k^{\prime\prime}}(c-1)\gamma(k^{\prime\prime},c)}
=\frac{\gamma(k,c)}{\sum_{k^{\prime\prime}}\gamma(k^{\prime\prime},c)},
\end{equation}
and the probability that the chosen $c$-clique edge links nodes of
degree $k$ and $\kp$ is
\begin{equation}
P_c(k,\kp) = s_k^{(c)} s_\kp^{(c)} \quad\text{  for } c>1.
\label{Pc}
\end{equation}
Inserting (\ref{Pe}) and (\ref{Pc}) into (\ref{new1}) enables us
to write
 the global $P(k,\kp)$ matrix for the network as
\begin{eqnarray}
P(k,\kp) &=& \alpha^{(1)} s_k^{(1)}s_\kp^{(1)} + \sum_{c>1}
\alpha^{(c)}
s_k^{(c)}s_\kp^{(c)}\nonumber\\
&=& \sum_{c \ge 1} \alpha^{(c)} s_k^{(c)} s_\kp^{(c)}.
\label{Pkkp}
\end{eqnarray}
We can then calculate $p_{th}^{(2)}$, the bond percolation
threshold in an unclustered network with the same degree-degree
correlations as the original network \cite{Newman02,Vazquez03}, as
$p_{th}^{(2)}=1/\lambda_{max}$, where $\lambda_{max}$ is the
largest eigenvalue of the matrix $\mathbf{C}$ with entries given
by
\begin{equation}
C_{k,j} = \frac{(j-1)}{\sum_\kp P(k,\kp)} P(k,j).
\end{equation}

Moreover, we can see that $\gamma$-theory networks are necessarily
assortative by showing that
\begin{equation}
\sum_{k,\kp} k P(k,\kp) \kp - \left( \sum_{k,\kp} k
P(k,\kp)\right)^2 \ge 0. \label{C10}
\end{equation}
This quantity determines the sign of the Pearson correlation
coefficient $r$ defined in Eq.~(3) of \cite{Newman02}, with
positive values corresponding to assortative networks. Using
(\ref{Pkkp}), the left-hand side of (\ref{C10}) may be written as
\begin{equation}
\sum_c \alpha^{(c)} x_c^2 - \left( \sum_c \alpha^{(c)} x_c
\right)^2,
\end{equation}
where $x_c = \sum_k k s_k^{(c)}$ and $\sum_c \alpha^{(c)} =1$, so
this expression may be rewritten as
\begin{equation}
\frac{1}{2}\sum_{c,c^\prime} \alpha^{(c)}\alpha^{(c^\prime)}
\left(x_c-x_{c^\prime}\right)^2.
\end{equation}
Since all $\alpha^{(c)}$ terms are non-negative the inequality
(\ref{C10}) must hold, and the $\gamma$-theory networks are
assortative.

We emphasize the fact that asortativity follows  here directly
from  the decomposition (\ref{Pkkp}) of $P(k,\kp)$ into disjoint
parts, each of which has the form of a randomly-connected network.
In Newman's recent clustering model \cite{Newman09a}, for example,
there are also two types of links: those which are edges of
triangles, and those which are not. Stubs of each of these two
types are randomly connected to stubs of the same type---it
follows that the $P(k,\kp)$ matrix for Newman's theory must be of
the general form (\ref{Pkkp}), and therefore networks generated by
his model must also be assortative.

\section{Example of correlation in colored-edge networks}
\label{newApp} We consider a particular example of the non-trivial
correlation structure of the colored-edge networks described in
Section~\ref{sect4} (and further analyzed in Appendix~\ref{appD}).
Consider a colored-edge network corresponding to the example
(\ref{gamma4}), where half the nodes are of degree $k=2$ and half
are of degree $k=3$. We choose parameters $\alpha=0$ and
$\beta=1$, which means that every $k=2$ node has two green stubs,
and every $k=3$ node has 1 green and 2 red stubs. Pairs of green
stubs are chosen at random to form green edges, and similarly for
red stubs/edges. The nearest-neighbor correlations are given by
the $P(k,\kp)$ matrix defined in (\ref{Pkkp}); for the parameters
chosen here we have $P(2,2) = 4/15 $, $P(2,3)=P(3,2)= 2/15 $, and
$P(3,3) =7/15 $.

Let us now consider degree correlations beyond nearest-neighbors.
Specifically, we choose a node of degree 3 and examine the
fraction of its second neighbors which are also of degree 3
(ignoring cycles in the $N\to \infty$ limit). We denote this
quantity $Q(3|3)$, as it is the probability that node $A$ has a
second neighbor of degree 3, given that node $A$ itself has degree
 3.

 Since the degree distribution of first-neighbors of $A$ is given exactly
 by
 \begin{equation}
 P(k|3) = \frac{P(k,3)}{\sum_{\kp}P(\kp,3)} \quad\text{ for }k=2,3,
 \end{equation}
 it is tempting to calculate second-neighbor correlations under
 the Markovian assumption that the network is completely described
 by its $P(k,\kp)$ distribution. This assumption underlies the
 calculation of the threshold we denote as $p_{th}^{(2)}$, and if
 applied to our example would estimate the value of $Q(3|3)$ by
 \begin{equation}
 \sum_{\kp}P(3|\kp)P(\kp|3) = \frac{55}{81} \label{Qest}.
 \end{equation}

 However, the coloring of the edges implies that the true $n$th-nearest-neighbor correlation
 structure is not adequately described by $P(k,\kp)$ for $n>1$. To
 show this, we now calculate the exact value of $Q(3|3)$ and show
 that it differs from the Markovian-assumption estimate
 (\ref{Qest}). First, note that since all $k=3$ nodes have 1 green
 stub (as well as 2 red stubs) and all $k=2$ nodes have 2 green
 stubs, travelling along a random green edge will lead to a $k=3$
 node with probability $1/3$, and to a $k=2$ node with probability
 $2/3$. Similarly, travelling along a random red edge leads to a
 $k=3$ node with probability 1.

 Let us start at the $k=3$ node called $A$, and enumerate all
 possible paths leading to degree-3 second neighbors of $A$, thus
 calculating $Q(3|3)$. A fraction
 $1/3$ of A's first neighbors are accessed via green edges,
 with the remaining fraction $2/3$ being accessed by travelling along a red
 edge. Suppose first that we travel along a green edge from $A$.
 With probability $1/3$ the green edge leads to a $k=3$ neighbor,
 otherwise the neighbor has $k=2$. If the neighbor has $k=3$, and
 noting that we arrived at him along a green edge, his connections
 to second neighbors of $A$ are necessarily along red edges, and so these second neighbors
 have degree $k=3$ with probability 1. On the other hand, if the
 first neighbor of $A$ has $k=2$, the access to $A$'s second
 neighbor along this path must be along a green edge, and so the second neighbor
 found on this path is of degree 3 with probability $1/3$.

 To summarize so far: starting from a $k=3$ node $A$ we can find
 degree-3 second neighbors of $A$ by proceeding
 \begin{itemize}
 \item along a green edge (prob $1/3$) via a $k=3$ first neighbor
 (prob $1/3$) and then along a red edge (prob 1). Total probability:
 $1/9$.
 \item or, along a green edge (prob 1/3) via a $k=2$ first neighbor
 (prob $2/3$) and then along a green edge (prob $1/3$). Total
 probability: $2/27$.
 \end{itemize}
 Similar arguments show that the remaining possible paths proceed
 from $A$
 \begin{itemize}
 \item along a red edge (prob $2/3$) via a first neighbor of
 degree-3 (prob 1) and then either along a red edge (prob $1/2$)
 to a $k=3$ node (prob 1), or along a green edge (prob $1/2$) to a
 $k=3$ node (prob $1/3$). Total probability: $4/9$.
 \end{itemize}
Summing over all possible paths we obtain
\begin{equation}
Q(3|3) = \frac{1}{9} + \frac{2}{27} + \frac{4}{9} = \frac{17}{27},
\end{equation}
which differs from the value $55/81$ obtained in (\ref{Qest})
under the Markovian approximation. We conclude that in
colored-edge networks (and hence in the $\gamma$-theory clustered
networks) $n$th-nearest-neighbor correlations beyond $n=1$ are not
completely described by the $P(k,\kp)$ distribution under the
Markovian assumption.

\section{Percolation in colored-edge networks} \label{appD}
We consider bond percolation in an unclustered network of $N$
nodes (in the $N \to \infty$ limit), composed of two types of
edges  (green or red) as described in Section~\ref{sect4}. Such
networks may be created by considering a $\gamma$-theory network
with only one non-trivial clique class $c=\overline c$ and with
the internal $c$-clique edges colored red while the external links
are colored green, see Fig.~\ref{fig4} for an example with
$\overline c =3$. A similar idea is used in \cite{Miller09} for
Newman's triangle-based networks \cite{Newman09a}. The total
number of green stubs (half-edges) is
\begin{eqnarray}
N \sum_{k,c}(k-c+1)\gamma(k,c) &=& N \sum_k k \gamma(k,1)
+\nonumber\\&&\hspace{-1.2cm}+ N \sum_k (k-\overline c
+1)\gamma(k,\overline c),
\end{eqnarray}
and the total number of red stubs is
\begin{equation}
N \sum_k (\overline c -1)\gamma(k,\overline c),
\end{equation}
since any node with red stubs has exactly $\overline c -1$ of
them. Green stubs are randomly linked to green stubs, and
similarly for red stubs. As in \cite{Gleeson09b,Gleeson09a}, we
define a node as active if it is part of the GCC, and assume all
nodes are initially inactive. Using a tree structure, define $q_g$
as the probability that a node with a green edge linking to its
parent is active, and $q_r$ is the corresponding probability for a
node with a red edge leading to its parent. Then standard
arguments (see, for example, \cite{Gleeson08a,Gleeson09b}) lead to
the following self-consistent equations for $q_g$ and $q_r$:
\begin{eqnarray}
q_g &=& G(q_g,q_r) \nonumber\\
q_r &=& R(q_g,q_r), \label{D3}
\end{eqnarray}
where the functions $G$ and $R$ are defined as
\begin{eqnarray}
G(q_g,q_r) &=& \sum_{k,c} \frac{(k-c+1) \gamma(k,c)}{z_e} \times
\nonumber\\
&&\hspace{-0.2cm}\times\left[ 1 - (1-p
q_g)^{k-c}(1-p q_r)^{c-1}\right], \label{qg} \\
R(q_g,q_r) &=& \sum_{k} \frac{ \gamma(k,\overline
c)}{\sum_\kp\gamma(\kp,\overline c)} \times \nonumber\\
&&\hspace{-0.2cm}\times \left[ 1 - (1-p q_g)^{k-\overline c+1}(1-p
q_r)^{\overline c-2}\right]. \label{qr}
\end{eqnarray}
Similarly, the final density of active nodes, i.e., the GCC size,
is given by
\begin{equation}
S = \sum_{k,c} \gamma(k,c) \left[ 1-(1-p q_g)^{k-c+1}(1-p
q_r)^{c-1}\right].
\end{equation}
The percolation threshold point is determined by standard cascade
condition arguments \cite{Gleeson08a} applied to the system
(\ref{D3})-(\ref{qr}). Defining $\mathbf{B}$ as the matrix
\begin{eqnarray}
\mathbf{B} &=& \frac{1}{p} \left. \left[
\begin{array}{cc}
   \frac{\partial G}{\partial q_g}   & \frac{\partial G}{\partial q_r}
   \\
    \frac{\partial R}{\partial q_g}  & \frac{\partial R
    }{\partial q_r}
\end{array}
\right] \right|_{q_g=q_r=0},
\end{eqnarray}
which has elements
\begin{eqnarray}
B_{11}&=& \frac{1}{z_e}
\sum_{k,c}(k-c+1)(k-c)\gamma(k,c)\nonumber\\
B_{12}&=&\frac{(\overline c -1)}{z_e}\sum_k (k-\overline c +1)
\gamma(k,\overline c)\nonumber\\
B_{21}&=&\frac{1}{\sum_\kp \gamma(\kp,\overline c)}    \sum_k
(k-\overline c +1)\gamma(k,\overline c) \nonumber\\
B_{22} &=& \overline c -2, \label{Bvals}
\end{eqnarray}
the percolation threshold is given by
$p_{th}^{(ce)}=1/\lambda_{max}$ where $\lambda_{max}$ is the
larger of the eigenvalues of $\mathbf{B}$, i.e.,
\begin{equation} p_{th}^{(ce)}=
\frac{2}{B_{11}+B_{22}+\sqrt{(B_{11}-B_{22})^2+4 B_{12}B_{21}}}.
\label{pce}\end{equation} Since all the $B_{ij}$ elements are
non-negative, we have the bound
\begin{equation}
p_{th}^{(ce)} \le \frac{1}{B_{22}} \label{Dnew}
\end{equation}
(this follows by noting $\sqrt{(B_{11}-B_{22})^2+4 B_{12}B_{21}}
\ge B_{22}-B_{11}$) which we will use below.

 Next we show that
$p_{th}^{(ce)}\le p_{th}^{(\gamma)}$ for networks of this type.
From Eq.~(\ref{7}), note that $p_{th}^{(\gamma)}$ is the solution
of the polynomial equation $H(p)=1$, where
\begin{eqnarray}
H(p) &=& \frac{1}{z_e} \sum_{k,c}
(k-c+1)\gamma(k,c)\times\nonumber\\
&&\times\left(p(k-c)+(z_c-c+1)D_c(p)\right) \nonumber\\
&=& B_{11} p +\frac{1}{\overline{c}-1}
B_{12}B_{21}D_{\overline{c}}(p),
\end{eqnarray}
and $B_{ij}$ refers to the entries of the non-negative matrix
$\mathbf{B}$ above. Following the arguments of
Appendix~\ref{appB}, we will show that
$H\left(p_{th}^{(ce)}\right) \le 1$ by using the bound (\ref{B2})
on $D_{\overline{c}}(p)$. This gives
\begin{equation}
H(p) \le B_{11} p + B_{12}B_{21}\frac{p^2}{1-p(\overline{c}-2)}
\label{D11}
\end{equation}
for all $p$ such that $0\le p \le \frac{1}{\overline{c}-2}$.

Noting that $B_{22}=\overline c - 2$, we see from (\ref{Dnew}) the
inequality  $p_{th}^{(ce)}\le 1/(\overline c -2)$ is obeyed and so
we may apply (\ref{D11}) with $p=p_{th}^{(ce)}$.  Substituting
$p=p_{th}^{(ce)}$ from (\ref{pce}) (with (\ref{Bvals})) into
(\ref{D11}) and simplifying yields
\begin{equation}
H\left(p_{th}^{(ce)}\right) \le 1,
\end{equation}
and the result $p_{th}^{(ce)}\le p_{th}^{(\gamma)}$  follows.

\bibliography{networks}
\end{document}